\documentclass[floatfix,twocolumn,aps,showpacs]{revtex4}

\usepackage{pstricks}
\usepackage{pst-coil}
\usepackage{amssymb}
\usepackage{graphics,epsf}

\begin{document}

\title{Is the shell-focusing singularity of Szekeres space-time visible?}
\author{Brien C Nolan\footnote{brien.nolan@dcu.ie}
} \affiliation{School of Mathematical Sciences, Dublin City
University, Glasnevin, Dublin 9, Ireland.}
\author{Ujjal Debnath\footnote{ujjal@iucaa.ernet.in, ujjaldebnath@yahoo.com}}
\affiliation{Department of Mathematics, Bengal Engineering and
Science University, Shibpur, Howrah-711 103, India.}
\begin{abstract}
The visibility of the shell-focusing singularity in Szekeres
space-time - which represents quasi-spherical dust collapse - has
been studied on numerous occasions in the context of the cosmic
censorship conjecture. The various results derived have assumed that
there exist radial null geodesics in the space-time. We show that
such geodesics do not exist in general, and so previous results on
the visibility of the singularity are not generally valid. More
precisely, we show that the existence of a radial geodesic in
Szekeres space-time implies that the space-time is axially
symmetric, with the geodesic along the polar direction (i.e.\ along
the axis of symmetry). If there is a second non-parallel radial 
geodesic, then the space-time is spherically symmetric, and so is a
Lema\^{\i}tre-Tolman-Bondi (LTB) space-time. For the case of the
polar geodesic in an axially symmetric Szekeres space-time, we give
conditions on the free functions (i.e.\ initial data) of the
space-time which lead to visibility of the singularity along this
direction. Likewise, we give a sufficient condition for censorship
of the singularity. We point out the complications involved in
addressing the question of visibility of the singularity both for
non-radial null geodesics in the axially symmetric case and in the
general (non-axially symmetric) case, and suggest a possible
approach.
\end{abstract}
\maketitle

\newtheorem{assume}{Assumption}
\newtheorem{theorem}{Theorem}
\newtheorem{prop}{Proposition}
\newtheorem{corr}{Corollary}
\newtheorem{lemma}{Lemma}
\newtheorem{remark}{Remark}
\newtheorem{definition}{Definition}
\newtheorem{remarkthe}{Remark}[theorem]
\newtheorem{remarkcorr}{Remark}[corr]
\newtheorem{remarklem}{Remark}[lemma]
\newcommand{\nn}{\nonumber}
\newcommand{\ds}{\displaystyle}
\newcommand{\pd}[2]{\frac{\partial #1}{\partial #2}}
\newcommand{\pdd}[3]{\frac{\partial^2 #1}{\partial #2\partial #3}}
\newcommand{\be}{\begin{equation}}
\newcommand{\ee}{\end{equation}}
\newcommand{\bq}{\begin{eqnarray}}
\newcommand{\eq}{\end{eqnarray}}
\newcommand{\by}{\begin{eqnarray*}}
\newcommand{\ey}{\end{eqnarray*}}
\newcommand{\bz}{\bar{\zeta}}
\newcommand{\bx}{\bar{\xi}}
\newcommand{\td}{\dot{t}}
\newcommand{\rd}{\dot{r}}
\newcommand{\xd}{\dot{x}}
\newcommand{\yd}{\dot{y}}
\newcommand{\so}{{\cal{O}}}
\newcommand{\ch}{{\cal{H}}}
\newcommand{\ce}{{\cal{E}}}
\newcommand{\cf}{{\cal{F}}}
\newcommand{\cm}{{\cal{M}}}
\newcommand{\cmt}{\tilde\cm}
\newcommand{\pnc}{{\cal{N}}}
\newcommand{\bphi}{{\tilde{\phi}}}
\newcommand{\hE}{{\hat{E}}}
\newcommand{\kp}{\kappa}
\def\fin{\hfill \rule{2.5mm}{2.5mm}\\ \vspace{0mm}}
\def\bs{\rule{2.5mm}{2.5mm}}
\section{Introduction and summary.}
The weak cosmic censorship hypothesis (CCH) maintains that
realistic gravitational collapse leads to the formation of a
black hole rather than a naked singularity. Among the different
studies of the hypothesis, we mention two categories of interest.
First, there are demonstrations of the validity of the hypothesis
in specific circumstances (the prime example of this is
Christodoulou's proof of the instability - and hence
non-realistic nature - of naked singularities in the spherical
collapse of a minimally coupled scalar field \cite{christo1}).
The other category involves the construction of an example of a
space-time which undergoes collapse from a regular configuration
to a naked singularity. Many such examples have been constructed,
but to date, none has been shown to involve both (i) a physically
realistic matter model and (ii) stability in the initial data
space of those space-times which give rise to naked
singularities. It is probably fair to say that the main utility
of the latter class of studies has been to refine and better
understand the content of the CCH.

These examples have mainly involved spherically symmetric
space-times, for example the shell-crossing \cite{yod1,yod2} and
shell-focusing \cite{christo2} singularities in
Lema\^{\i}tre-Tolman-Bondi (LTB) spherical dust collapse, and the
naked singularity solutions that arise at the threshold of black
hole formation in scalar field and perfect fluid collapse
\cite{gundlach1}. A notable (although as we will see flawed)
exception to this involves the various studies of the visibility
of the shell-focusing singularity in Szekeres space-time
\cite{joshi-krolak}-\cite{debnath4}. First analyzed by Szekeres in
\cite{szekeres1}, this class of space-times corresponds to
solutions of the Einstein equations for dust, where the fluid
flow vector is geodesic and non-rotating. The metric admits no
Killing vector fields (in the general case) but for reasons
described below is referred to as quasi-spherical. It can be
understood as a non-spherical generalization of the LTB class of
dust-filled space-times, and its evolutionary aspects are very
closely related to those of the corresponding LTB models, and so
are relatively straightforward. Hence analyzing the visibility or
otherwise of singularities that arise in this model affords an
opportunity to study cosmic censorship in non-spherical collapse.

Szekeres initiated the study of the singularities in this model,
and noted the possibility that the shell-crossing singularity (see
below) may be visible \cite{szekeres2}. He also noted the
occurrence of an apparent horizon (or in current terminology, a
marginally trapped tube) that forms at least as early as the
shell-focusing singularity (again, see below). As in the spherical
case, the shell-crossing singularity is interpreted as being
fundamentally non-gravitational in origin and not of particular
significance for cosmic censorship. Thus attention turned to the
shell-focusing singularity, where in the spherical case, a
more-or-less complete understanding of the visibility or otherwise
of the singularity has been developed. In these studies
\cite{joshi-krolak}-\cite{debnath4}, the question of visibility of
the shell-focusing singularity is considered from the point of
view of the existence or otherwise of future-pointing null
geodesics with an additional simplifying property that allows one
to refer to these geodesics as `radial' (see below). We show below
that {\em radial geodesics do not exist in a general Szekeres
space-time} and consequently, {\em the analyses of the question of
visibility of the singularity of
\cite{joshi-krolak}-\cite{debnath4} are not generally valid}.
Motivated by this observation, we revisit the question of the
visibility of the shell-focusing singularity. We provide some
preliminary results on this question, and in particular, consider
it in the case when the space-time is axially symmetric. In this
case, a single radial geodesic direction exists, and the analysis
of the visibility of the singularity along this direction is
essentially the same as the spherically symmetric case. We
emphasize that the question becomes considerably more difficult in
the general (non-axially symmetric) case.

In the next section, we review the basic properties of Szekeres
space-time, and discuss what is meant by referring to this
space-time as `quasi-spherical'. We then discuss the formation of
singularities in a collapsing Szekeres space-time, and discuss the
conditions on the free metric functions that arise from the
imposition of regularity conditions on the initial data. Here, we
specialize to the so-called marginally bound case. We will
indicate clearly when this restriction is in place, and when
results apply generally.

In section 3, we analyze the geodesic equations of Szekeres
space-time and show that the existence of a radial geodesic
implies that the space-time is axially symmetric. Furthermore,we
show that the existence of a second non-collinear radial geodesic
implies that the space-time is spherically symmetric. The
marginally bound assumption is {\em not} required in this section.

In section 4, we derive some elementary results in the marginally
bound case relating to the visibility of the shell-focusing
singularity, which is spherically symmetric in the sense that it
corresponds to a surface $t=t_c(r), r\geq0$. This singularity is
always preceded by an apparent horizon given by $t=t_{ah}(r)$: we
have $t_{ah}(r)\leq t_c(r)$ for all $r$ with equality if and only
if $r=0$. The region of space-time with $t_{ah}(r)<t<t_c(r)$ is
trapped. Then intuitively, one expects that only the central
singularity $(t,r)=(t_c(0),0)$ can be visible. This is immediate
in spherical symmetry, but is slightly non-trivial in the
quasi-spherical case - the result is confirmed nonetheless. We
derive the related result that a geodesic that emerges into the
future from the central singularity must emerge into the untrapped
region $t<t_{ah}$, and hence show that the singularity is censored
if $t_{ah}^\prime(r)<0$ (this condition can be described in terms
of the initial data of the collapse). We also show that for
sufficiently small values of $r$, the apparent horizon is a
one-way membrane for all future-pointing causal geodesics: such
geodesics cannot leave the trapped region. Again, this is
immediate in the spherical case, but requires checking in the
quasi-spherical case.

In section 5, again working in the marginally bound case, we
consider the question of the visibility of the singularity for the
polar null geodesic of the axially symmetric models. This problem is
essentially the same as the corresponding spherically symmetric
problem, and we can give sufficient conditions (in terms of the
initial data) for the formation of a naked singularity.

We conclude by briefly considering the substantive, and crucially,
open, question of the visibility of the shell-focusing
singularity in quasi-spherical collapse. We point out how this
question is considerably more difficult than in the spherical case
and suggest an approach to its consideration. We set $8\pi G=c=1$.

\section{Szekeres space-time and its singularities.}

A comprehensive review of the properties of the Szekeres
space-times representing non-accelerating, irrotational dust is
given in \cite{kras}; this review includes equivalent invariant
characterizations of the class which involve some technicalities
that will not play any role here. Suffice to say that these lead
uniquely to the line element \be
ds^2=-dt^2+e^{2\alpha}dr^2+e^{2\beta}(dx^2+dy^2)\label{sz-lel}\ee
where \begin{eqnarray} e^\beta &=&R(t,r)e^\nu,\label{bedef}\\
e^{-\nu}&=&A(r)(x^2+y^2)+B_1(r)x+B_2(r)y+C(r),\label{nudef}\\
e^\alpha
&=&\frac{R^\prime+R\nu^\prime}{\sqrt{1+f(r)}},\label{aldef}
\end{eqnarray}
where the prime denotes differentiation with respect to $r$. The
geodesic fluid flow vector is $\ds{\frac{\partial}{\partial t}}$ and
the coordinates $(r,x,y)$ are co-moving. The ranges of these
coordinates are $(x,y)\in \mathbb{R}^2$, $r\geq0$ and that of $t$
will be discussed below. The free functions $A, B_1, B_2$ and $C$
are related by \be B_1^2+B_2^2-4AC=-1.\label{norm}\ee (There are
also solutions with $0$ and $+1$ on the right hand side here: the
choice $-1$ forms part of the input essential to the interpretation
of these space-times as being quasi-spherical.) The remaining
Einstein equations determine the evolution of $R$ and define the
energy density of the space-time: \be \left(\pd{R}{t}\right)^2 =
f(r)+\frac{F(r)}{R},\label{r-evolve}\ee \be
\rho(t,r,x,y)=\frac{F^\prime+3F\nu^\prime}{R^2(R^\prime+R\nu^\prime)}.\label{rho-def}\ee
$F\geq0$ is a function of integration. We note that the coordinate
freedom corresponds to rescalings of the co-moving radial coordinate
$r\to \hat{r}(r)$ (which must be a monotone mapping), the
transformations in the $(x,y)-$plane discussed in (\ref{flt}) below
and trivial shifts of the origin of $t$.

The quasi-spherical interpretation arises as follows
\cite{szekeres2}: each 2-surface $S_{t,r}$ of constant $t$ and $r$
is a round 2-sphere with proper radius $R(t,r)$. To see this, we
write the line element $dl^2_{(t,r)}$ of $S_{t,r}$ in the
stereographic coordinates $\zeta = x+iy$: \[ dl^2_{(t,r)}
=\frac{R^2(t,r)}{(A(r)\zeta{\bar{\zeta}}+ B(r)\zeta+\bar{B}(r)\bz
+ C(r))^2}d\zeta d\bz\] where $B=(B_1-iB_2)/2$. The form of this
line element is invariant under the fractional linear
transformation \be \zeta \to \xi= \frac{k\zeta+l}{m\zeta+n},\quad
kn-lm=1\label{flt}\ee and for each fixed value of $t$ and $r$,
such a transformation can be found so that the line element has
the form
\[ dl^2_{(t,r)}=R^2(t,r)\frac{4d\xi d\bx}{(1+\xi\bx)^2},\]
which is the line element of the round 2-sphere with radius $R$. It
should be noted that the condition (\ref{norm}) plays a crucial role
in deriving the explicitly spherical form of $dl^2_{(t,r)}$. The
transformation $(x,y)\to(\theta,\phi)$ given by
\[ \zeta = e^{i\phi}\cot\frac{\theta}{2}\] yields the more
familiar spherical form
\[ dl^2_{(t,r)} = R^2(t,r)(d\theta^2+\sin^2\theta d\phi^2).\]
The fact that a different transformation (\ref{flt}) is required
for each different $S_{t,r}$ indicates that in each 3-space of
constant $t$, these 2-spheres are not concentric.

It will be useful to consider the form of the 4-dimensional line
element using the spherical coordinates $(\theta,\phi)$. This
yields \be ds^2= -dt^2
+e^{2\alpha}dr^2+e^{2\beta}(d\theta^2+\sin^2\theta
d\phi^2),\label{lel-spherical}\ee where now \bq
e^\beta&=&R(t,r)e^\mu,\label{be-def-sph}\\
e^{-\mu} &=&
a\cos\theta+b_1\sin\theta\cos\phi+b_2\sin\theta\sin\phi+c,\label{mu-def}\\
e^\alpha&=&\frac{R^\prime+R\mu^\prime}{\sqrt{1+f}}.\label{al-def-sph}\eq
The functions $a, b_i, c$ are related to $A, B_i, C$ by \be
a=A-C,\quad b_i=B_i,\quad c=A+C.\label{rels}\ee The condition
(\ref{norm}) reads \be c^2-a^2-b_1^2-b_2^2=1.\label{sphnorm}\ee

Before proceeding to discuss gravitational collapse and the
formation and nature of singularities in these space-times, we
note that the spherical limit arises when and only when
$B_1=B_2=0$ and $A$ and $C$ are constant and equal (and so by
(\ref{norm}) both equal to 1/2). Furthermore, it is clear from
(\ref{lel-spherical}) and (\ref{mu-def}) that the space-time is
axially symmetric with Killing vector
$\ds{\frac{\partial}{\partial \phi}}$ when $B_1=B_2=0$.

In order to model gravitational collapse using a Szekeres
space-time, we choose the negative root of (\ref{r-evolve}). The
resulting equation, and the analysis of its consequences, are
greatly simplified by taking $f(r)=0$. By analogy with the
spherical case, this is referred to as the marginally bound case.
The equation is then easily integrated and the solution can be
written in the form \be
R^3=\frac{9}{4}F(t_c(r)-t)^2,\label{Rsol}\ee where $t_c$ is a
function of integration that describes the time at which the
2-sphere $S_{t,r}$ collapses to zero radius. This is called the
shell-focusing singularity: all the `shells' $S_{t,r}$ collapse
to zero radius at this surface.

It is convenient to exploit the freedom in the co-moving radial
coordinate $r$ to set $R=r$ on an initial surface: by an allowed
shift of the origin of $t$, we can take this to be $t=0$ without
loss of generality. Thus $R(0,r)=r$ and hence $R^\prime(0,r)=1$.
With this choice, we have \be
t_c(r)=\frac23\sqrt{\frac{r^3}{F}},\label{tcdef}\ee which leads to
the following convenient form of (\ref{Rsol}): \be
\left(\frac{R}{r}\right)^3=\left(1-\frac{t}{t_c}\right)^2.\label{Rsol1}\ee

It can be shown that the Kretschmann scalar of (\ref{sz-lel})
diverges if and only if the density (\ref{rho-def}) does so. Thus
scalar curvature singularities can be discussed in terms of this
quantity alone. As well as the shell-focusing singularity at
$t=t_c(r)$ for which $R=0$, there may be singularities when
$R^\prime+R\nu^\prime$ vanishes. By analogy with the spherical
case, these are referred to as shell-crossing singularities. (The
analogy is perhaps not quite appropriate: shells of Szekeres
space-time, i.e.\ the 2-spheres $S_{t,r}$ will cross if we
encounter $R(t,r_1)=R(t,r_2)$ for some $r_1\neq r_2$ and some
$t>0$. A necessary and sufficient condition for this to occur is
$R^\prime(t,r)=0$ at some $t>0,r>0$.) A crucial task is to rule
out the occurrence of both types of singularity on the initial
slice. To this end, we note that the initial density is given by
\[ \rho_0(r):=\rho(0,r) =
\frac{F^\prime+3F\nu^\prime}{r^2(1+r\nu^\prime)}.\] We will
require this term to be non-negative and finite for all $r\geq 0$.
Thus we impose \be F^\prime+3F\nu^\prime\geq 0,\label{f-con1}\ee
\be 1+r\nu^\prime>0 \label{nu-con1}\ee for all $r\geq0$ and all
$(x,y)\in\mathbb{R}^2$. We note that we also impose $e^{-\nu}>0$.
Identical conditions hold with $\nu$ replaced by $\mu$. These
conditions - i.e.\ (\ref{nu-con1}) and $e^{-\mu}>0$ - have been
considered by Szekeres \cite{szekeres2}, and it is worth repeating
the result here as we will use the corresponding conditions on
$a,b_i$ and $c$ below:

\begin{lemma}
\noindent (i) Let the condition (\ref{norm}) hold. Then
$e^{-\nu}=A(x^2+y^2)+B_1x+B_2y+C$ is positive  for all
$(x,y)\in\mathbb{R}^2$ and all $r\geq0$ if and only if \be A(r)>0
\quad {\hbox{for all}}\quad r\geq0.\label{a-con1}\ee  (ii) Assuming
the conditions (\ref{norm}) and $e^{-\nu}>0$,
\[1+r\nu^\prime>0\] for all $(x,y)\in\mathbb{R}^2$ and all $r\geq0$
if and only if \be A-rA^\prime>0  \quad{\hbox{for all}}\quad
r\geq0\label{a-con2}\ee and \be A^\prime C^\prime -
B^\prime{\bar{B}}^\prime>-\frac{4}{r^2}\quad{\hbox{for all}}\quad
r\geq0\label{a-con3}\ee \hfill$\blacksquare$
\end{lemma}

Requiring that $\rho_0$ be finite in the limit as $r\to0$,
Szekeres \cite{szekeres2} also derives the condition \be
F(r)=O(r^3),\quad r\to 0 \label{Fcon1}\ee which we shall assume
henceforth.

Next, we will consider conditions that rule out the occurrence of
a shell-crossing singularity ($R^\prime+R\nu^\prime=0$) prior to
the occurrence of the shell-focusing singularity. That is, we
seek conditions on the initial data functions so that
\[ R^\prime+R\nu^\prime>0, \quad{\hbox{for all}}\quad 0\leq t <t_c(r),
r\geq0.\] We note that this condition follows by (\ref{nu-con1})
{\em if} the inequality $R^\prime-R/r\geq0$ holds. But using
(\ref{Rsol1}), this latter condition is equivalent to
$t_c^\prime\geq0$. Hence the condition \be F^\prime -
3\frac{F}{r}\leq 0,\quad r\geq 0,\label{Fcon2}\ee which is
equivalent to $t_c^\prime\geq 0$, is a condition that can be
imposed on the initial data and that guarantees the absence of
shell-crossing singularities. It is worth noting that for $t>0$,
(\ref{Fcon2}) is {\em equivalent} to the condition \be
R^\prime\geq\frac{R}{r}.\label{Rcon}\ee In a sense, this is the
best bound that can be imposed. First, it includes all cases of
interest: $t_c^\prime\geq 0$ is a necessary condition for the
visibility of the shell-focusing singularity. (To see this, we
simply note that $t$ must increase along a future pointing causal
geodesic emerging from the singularity.) Second, if the bound
(\ref{Fcon2}) is violated, then we can find examples of $\nu$
(which is constructed entirely from initial data functions) for
which the corresponding space-time will contain shell-crossing
singularities in its evolution.

To summarize, we assume the conditions on $A,B_i$ and $C$ of Lemma
1 and the conditions (\ref{Fcon1}) and (\ref{Fcon2}) on $F$. These
guarantee that the collapse proceeds from a regular state to the
shell-focusing singularity, and that no shell-crossing
singularities occur prior to the shell-focusing singularity.

\section{Geodesics, radial geodesics and symmetry}

With a clear description of the shell-focusing singularity in
place, we can now consider its causal nature or more accurately,
the question of its visibility. Of course this requires the
analysis of geodesics, and the determination of whether or not
there are future pointing causal geodesics that emerge from the
singularity. We will use a subscript ($\alpha_t$ etc.) to denote
partial derivatives with respect to $t,x$ and $y$. We retain the
prime for derivatives with respect to $r$, and an overdot will
represent differentiation along the geodesic (e.g.\ with respect
to an affine parameter for null geodesics). We also note that we
can drop the assumption that the space-time is marginally bound.
The geodesic equations for the line element (\ref{sz-lel}) are \bq
\ddot{t}+\alpha_te^{2\alpha}{\dot{r}}^2+\beta_te^{2\beta}({\dot{x}}^2+{\dot{y}}^2)&=&0,\label{geo-t}\\
\ddot{r}+\alpha^\prime\rd^2+2\alpha_t\td\rd+2\alpha_x\xd\rd+2\alpha_y\yd\rd&&\nn\\-\beta^\prime e^{2\beta-2\alpha}(\xd^2+\yd^2)&=&0,\label{geo-r}\\
\ddot{x}+\beta_x(\xd^2-\yd^2)+2\beta_y\xd\yd+2\beta_t\td\xd+2\beta^\prime\rd\xd&&\nn\\-\alpha_xe^{2\alpha-2\beta}\rd^2&=&0,\label{geo-x}\\
\ddot{y}+\beta_y(\yd^2-\xd^2)+2\beta_x\xd\yd+2\beta_t\td\yd+2\beta^\prime\rd\yd&&\nn\\-\alpha_ye^{2\alpha-2\beta}\rd^2&=&0,\label{geo-y}
\eq and we have the first integral \be
-\td^2+e^{2\alpha}\rd^2+e^{2\beta}(\xd^2+\yd^2)=\epsilon,\label{geo-int}\ee
where $\epsilon=0,+1,-1$ for null, space-like and time-like
geodesics respectively.

In \cite{joshi-krolak}-\cite{debnath4}, radial geodesics are defined
to be those along which $x$ and $y$ have constant values. From
(\ref{geo-x}) and (\ref{geo-y}), we see that along such a geodesic,
we must have \be
\pd{\alpha}{x}e^{2\alpha-2\beta}\rd^2=\pd{\alpha}{y}e^{2\alpha-2\beta}\rd^2=0.\label{rad-con}\ee
We rule out $e^{\alpha-\beta}=0$, as this corresponds to a
singularity (and the geodesics must reside in the space-time rather
than on its singular boundary). Both equations of (\ref{rad-con})
are satisfied if we take $\rd=0$. The only possible solutions of the
geodesic equations then have $\td^2=1=-\epsilon$: these are the
fluid flow lines and we will refer to these as trivial radial
geodesics. In particular, there are no {\em null} geodesics that
satisfy the condition $\rd=0$. The only possibility that remains in
(\ref{rad-con}) is that
\[ \pd{\alpha}{x}=\pd{\alpha}{y}=0\]
along the geodesic. It is clear that this is a restriction on the
metric functions. We can determine the exact geometric nature of
this restriction.

In order to do so, we note first that
\[
\pd{\alpha}{x}=\frac{Re^{-\alpha}}{\sqrt{1+f}}\pdd{\nu}{r}{x},\]
with a similar result holding for $\alpha_y$. Then a
straightforward calculation shows that the vanishing of $\alpha_x$
is equivalent to the vanishing of \bq Q_1&:=&
(AB_1^\prime-A^\prime B_1)(x^2-y^2)+2(AB_2^\prime-A^\prime
B_2)xy\nonumber\\&&+2(AC^\prime-A^\prime
C)x+(B_1B_2^\prime-B_1^\prime
B_2)y\nonumber\\&&+(B_1C^\prime-B_1^\prime C),\label{q1def}\eq
while vanishing of $\alpha_y$ is equivalent to vanishing of \bq
Q_2&:=& (AB_2^\prime-A^\prime B_2)(y^2-x^2)+2(AB_1^\prime-A^\prime
B_1)xy\nonumber\\&&+2(AC^\prime-A^\prime
C)y+(B_2B_1^\prime-B_2^\prime
B_1)x\nonumber\\&&+(B_2C^\prime-B_2^\prime C).\label{q2def}\eq

We can now prove the following result.

\begin{prop} If a Szekeres space-time admits a non-trivial radial geodesic,
then it also admits a Killing vector field generating an axial
isometry. \end{prop}

\noindent{\bf Proof:} Suppose that there is a non-trivial radial
geodesic along which $(x,y)=(x_0,y_0)$ is constant. Using a
transformation of the form (\ref{flt}), \[ \zeta=x+iy\to \xi =
u+iv = \frac{k\zeta+l}{m\zeta+n}\] we can assume without loss of
generality that $(x_0,y_0)=(0,0)$. The necessary conditions
$Q_1=Q_2=0$ for the existence of a non-trivial radial geodesic
then yield
\[ B_1=\lambda_1C,\quad B_2=\lambda_2 C \]
for some constants $\lambda_1, \lambda_2$. (We note that
(\ref{norm}) implies that $C\neq 0$.) If we consider a further
coordinate transformation of the form (\ref{flt}) - but with $l=0$
to preserve the origin - we find that (with $B=(B_1-iB_2)/2)$)
\[ B\to B_* = k\bar{n}B+m\bar{n}C = \bar{n}(k\lambda+m)C\]
where $\lambda=(\lambda_1-i\lambda_2)/2$. Since $kn-lm=kn=1$, we
have $k\neq0\neq\bar{n}$, and so we can choose $m=-k\lambda$ to
get $B_*=0$. Thus by using the coordinate freedom in the
stereographic coordinates $(x,y)$, we can write the line element
(\ref{sz-lel}) in a form in which $B_1=B_2=0$. As seen in Section
2 above, this is a sufficient condition for the space-time to be
axially symmetric, with axial Killing field given by
$\pd{}{\phi}=-y\pd{}{x}+x\pd{}{y}$.\hfill$\blacksquare$

\begin{remark}
{\em In this Proposition, the radial geodesic is normal to each of
the 2-spheres and emerges from the point with stereographic
coordinates $x=y=0$. In spherical coordinates, this corresponds to
$\theta=\pi$: the south pole in the standard configuration. The
north pole ($\theta=0$) corresponds to the point at infinity in
stereographic coordinates, and so is not covered by the coordinate
patch $(x,y)\in\mathbb{R}^2$. However this point can be included by
using an additional coordinate patch and it is then clear that there
is also a radial geodesic emerging from the north pole. We will
refer the these collinear radial geodesics as the polar geodesics.}
\end{remark}

\begin{corr}
If a Szekeres space-time admits two non-collinear non-trivial
radial geodesics, then the space-time is spherically symmetric.
\end{corr}
\noindent{\bf Proof:} From the previous proposition, we may
situate the first non-trivial geodesic in the direction
$(x,y)=(0,0)$. Then as we have seen $B_1=B_2=0$ and the space-time
is axially symmetric. A second non-collinear non-trivial radial
geodesic has $(x,y)=(x_0,y_0)\neq(0,0)$ constant along the
geodesic, and as we have seen, a necessary condition for the
existence of such a geodesic is that $Q_1$ and $Q_2$ vanish along
the geodesic. As $B_1=B_2=0$ and $x_0$ and $y_0$ are not both
zero, (\ref{q1def}) and (\ref{q2def}) yield
\[ AC^\prime-A^\prime C=0.\]
The condition (\ref{norm}) in the present case gives $AC=1/4$.
Combining this with the previous relation shows that $A$ and $C$ are
both constant. A transformation of the form (\ref{flt}) (with $l=0$
to preserve the origin) can then be used to set $2A=2C=1$, and so
the line element is spherically symmetric.\hfill$\blacksquare$

\begin{remark}
{\em We can give a geometric interpretation of this corollary.
Proposition 1 shows that to each non-trivial radial geodesic of
Szekeres there corresponds an axis of symmetry. Then the existence
of a second non-collinear non-trivial radial geodesic implies that
the space-time admits two non-parallel axes of symmetry and so
must be spherically symmetric.}
\end{remark}

\begin{remark}
{\em The results of this section imply that the analysis of
\cite{joshi-krolak}-\cite{debnath4}, which were carried out for
radial null geodesics, can only be valid for the polar geodesic of
an axially symmetric Szekeres space-time, or for a spherically
symmetric Szekeres space-time - i.e.\ LTB space-time. Thus the
question of the visibility of the shell-focusing singularity in a
general Szekeres space-time remains open. It is worth noting that
the imposition of the assumption of axial symmetry, equivalent to
setting $B_1=B_2$, implies that the number of free functions
$(A,B_1,B_2,f,F)$ has been reduced from 5 to 3. Therefore this
sector of the Szekeres class is highly specialized -  one could
impose a topology on the space of free functions in which the
non-axially symmetric solutions comprise an open dense subset of
the whole space - and so the results pertaining to the polar
geodesics cannot be assumed to reflect the general behaviour. It
is possible that the naked singularities found in the axially
symmetric case along the polar direction are (i) not visible from
any other direction and/or (ii) are not present in the
non-symmetric case.}
\end{remark}

\section{Some basic results in the marginally bound case.}
 We have argued above that the question of cosmic censorship in
 Szekeres is open. In the remainder of this paper, we seek to
 address this question. In this section, we will derive some
 results, valid in the marginally bound case only, which provide
 some useful preliminary results for the study of the visibility of
 the shell-focusing singularity. These results share the feature
 that they are trivial in the spherically symmetric case: in the
 quasi-spherical case, some checking is required.

 To begin, we recall from the work of Szekeres \cite{szekeres2} that an apparent
 horizon forms at the hypersurface $R(t,r)=F(r)$. That is, the outgoing future-pointing null geodesic
 normals to each 2-sphere $S_{t,r}$ have zero expansion on this
 hypersurface. By outgoing, we mean that $r$ increases with the
 affine parameter along the geodesic. Note also that while these
 geodesics are initially and instantaneously radial, this condition
 is immediately violated as the geodesic moves away from the
 2-sphere to which is was normal - see (\ref{geo-x}) and
 (\ref{geo-y}).

 From (\ref{Rsol1}), we can show that the apparent horizon is given
 by \be t=t_{ah}(r) = t_c(r) -\frac23F(r).\label{ah-def}\ee
 Thus the apparent horizon precedes the shell-focusing singularity
 ($F>0$ for $r>0$), and the condition (\ref{Fcon2}) implies that
 the apparent horizon and the shell-focusing singularity meet at
 the central singularity $(t,r)=(t_c(0),0)$. As shown in \cite{szekeres2},
 the region $t_{ah}(r)<t<t_c(r)$, $r\ge0$
 is trapped. That is, the 2-spheres $S_{t,r}$ are closed trapped
 surfaces in this region. The region $t<t_{ah}(r), r\geq0$ is
 untrapped.

 First, we point out that the portion $r>0$ of the shell-focusing
 singularity is censored.

 \begin{prop}
 There are no future-pointing causal geodesics of a marginally bound
 Szekeres space-time with past endpoint on the surface $t=t_c(r)$
 for any $r>0$.
 \end{prop}
 \noindent{\bf Proof:} We consider a future-pointing $(\td>0)$
 outgoing $(\rd>0)$ causal geodesic of (\ref{sz-lel}). Either the
 geodesic remains outgoing, or we encounter a value $\tau_0$ of the
 geodesic parameter $\tau$ (affine parameter or proper time) for which
 $\rd(\tau_0)=0$. But then (\ref{geo-r}) shows that this is a local
 minimum $(\ddot{r}>0)$ of $r$ along the geodesic. (To see this, we note that
 \[ \pd{\beta}{r}=(R^\prime+R\nu^\prime)e^\nu,\]
 which is positive by the no-shell crossing singularity condition.) As all stationary
 points must be local minima, there can in fact be only one local
 minimum. Thus $\rd(\tau)<0$ for all $\tau<\tau_0$. Suppose that
 this geodesic were to meet the singularity $t=t_c(r)$. Now along
 the geodesic we have
 \be
 \dot{R}=R_t\td+R^\prime\rd=-\sqrt{\frac{F}{R}}\td+R^\prime\rd.\label{R-geo}\ee
 Since $\td>0, R^\prime\geq 0$ and $\rd<0$, this must by negative in
 the approach to the singularity. But the singularity occurs at
 $R=0$, and so $R$ cannot increase into the past ($\dot{R}<0$) to reach the
 singularity, and so we get a contradiction.

We can now assume that $\rd>0$ for all $\tau\leq\tau_0$ where
$\tau_0$ is some arbitrary initial value for $\tau$. This being
the case, we can use $r$ as a parameter along the geodesic. Then
along the geodesic we may use (\ref{geo-int}) to write \be
\frac{dt}{dr}=e^\alpha\left[1-\frac{\epsilon}{\rd^2}e^{-2\alpha}+e^{2\beta-2\alpha}\left(\left(\frac{dx}{dr}\right)^2
+\left(\frac{dy}{dr}\right)^2\right)\right]^{\frac12}.\label{t-r}\ee
The fact that the geodesic is both future pointing and outgoing
indicates that the correct (positive) root has been taken here.
The change of $R$ along a future pointing geodesic that emerges
from the shell-focusing singularity at some $r>0$ satisfies the
following: \bq
\frac{dR}{dr}&=&R_t\frac{dt}{dr}+R^\prime\nn\\
&=&-\sqrt{\frac{F}{R}}\frac{dt}{dr}+R^\prime\nn\\
&<&-\frac{dt}{dr}+R^\prime\nn\\
&<&-e^\alpha+R^\prime = -R\nu^\prime\nn\\
&<&\frac{R}{r}.\label{rgeo-ineq}\eq The second line comes from the
field equation (\ref{r-evolve}) in the marginally bound case. The
third line arises due to the fact that for sufficiently small $R$,
the geodesic must be in the trapped region for which $F>R$. The
fourth line follows from (\ref{t-r}) above and the definition
(\ref{aldef}). The last line comes from the initial regularity
condition (\ref{nu-con1}). Integrating the overall inequality
proves the stated result: $R$ cannot reach zero (its value on the
shell-focusing singularity) unless $r$ also drops to
zero.\hfill$\blacksquare$

\begin{corr}
A future pointing causal geodesic with past endpoint on the
shell-focusing singularity must have its past endpoint on the {\em
central singularity} $(t,r)=(t_c(0),0)$. \hfill$\blacksquare$
\end{corr}

\begin{corr}
A future pointing null geodesic with past endpoint on the central
singularity must emerge into the untrapped region of space-time.
\end{corr}
\noindent{\bf Proof:} Suppose on the contrary that the geodesic
emerges into the trapped region, i.e.\  there exists $\delta>0$
such that $R(t(r),r)<F(r)$ for values of $R$ along the geodesic
and for all $0<r<\delta$. Letting $r_0\in(0,\delta)$ and
integrating (\ref{rgeo-ineq}) from $r$ to $r_0$ shows that in the
$R$-$r$ plane, $R$ stays above the line $R=\frac{R_0}{r_0}r$ for
all $r\leq r_0$. Using (\ref{Fcon1}), this implies that $R>F$ for
sufficiently small values of $r$, yielding a
contradiction.\hfill$\blacksquare$

\begin{prop}
If $t_{ah}^\prime(r)<0$ on $[0,\delta)$ for some $\delta>0$, then
the central singularity is censored.
\end{prop}
\noindent{\bf Proof:} The proof follows immediately from Corollary
3: if a geodesic were to emerge from the central singularity, then
it must emerge into the untrapped region and must have $t^\prime(r)$
non-negative for all sufficiently small values of $r$. This cannot
happen if $t_{ah}^\prime$ is negative in a neighbourhood of the
singularity.\hfill$\blacksquare$

\begin{remark}
{\em We note that this repeats in the Szekeres case a result that
holds in some generality in spherical symmetry \cite{magli1}, and
provides a sufficient condition, in terms of initial data, for the
singularity to be censored.}
\end{remark}

Finally in this section, we prove another result that mirrors
precisely the situation in the spherical case. As in that case, the
proof relies crucially on some of our assumptions about the
regularity of the initial data.

\begin{prop}
For sufficiently small values of $r$, the apparent horizon
$t=t_{ah}(r)$ acts as a one way membrane: a future pointing null
geodesic cannot cross the horizon from the trapped to the untrapped
region.
\end{prop}
\noindent{\bf Proof:} Let $p$ be a point of space-time on the
apparent horizon with $r>0$ and consider a future-pointing null
geodesic at $p$. If $\rd|_p=0$, then the fact that
\[ t_{ah}^\prime(r)=t_c^\prime(r)-\frac23F^\prime(r)\]
is finite for all $r>0$ and that $\td|p>0$ proves the stated result.
Suppose then that $\rd|_p\neq0$. Then there is a neighbourhood $I\ni
s_0$ such that $\rd(s)\neq 0$ for all $s\in I$ with $s|_p=s_0$ where
$s$ is the parameter along the geodesic in question. Then for points
on the geodesic corresponding to $I$, we can take $r$ to be the
parameter along the geodesic. Along a future pointing outgoing null
geodesic, we then have, using (\ref{t-r}),
\bq
\frac{dt}{dr}&>&e^\alpha\nn\\
&>&R^\prime-\frac{R}{r}\label{tr-bound}\eq where we have used
(\ref{aldef}) and the no-shell crossing condition (\ref{Rcon}).
Using (\ref{Rsol1}), we can show that \be
\left.\left(R^\prime-\frac{R}{r}\right)\right|_{t=t_{ah}(r)}=\frac{t_{ah}}{t_c}t_c^\prime.\label{t-r-on-ah}\ee
Then using $t_{ah}^\prime=t_c^\prime-\frac23F^\prime$, we can show
that
\[\left.\left(R^\prime-\frac{R}{r}\right)\right|_{t=t_{ah}(r)}>t_{ah}^\prime
\Leftrightarrow F^\prime-\frac{F}{r}>0.\] The initial regularity
condition (\ref{Fcon1}) indicates that this last inequality holds
for sufficiently small values of $r$. Thus when projected onto the
$r$-$t$ plane, for sufficiently small values of $r$, an outgoing
null geodesic can only cross the apparent horizon from below. The
same result is immediate for ingoing null geodesics as we can
show that subject to (\ref{Fcon1}), the slope of the apparent
horizon $t_{ah}^\prime$ is positive for small values of
$r$.\hfill$\blacksquare$

\begin{remark}
{\em We note that the result above is equivalent to stating that the
apparent horizon is space-like for small values of $r$. Globally,
there is no restriction: the horizon may be null or time-like for
larger values of $r$, and can change character. Thus the apparent
horizon is not always a one-way membrane in Szekeres space-time.}
\end{remark}

\section{Polar geodesics in the axially symmetric case}
As we have seen in Section 3, the only case in which radial
geodesics exist in Szekeres space-time is when the space-time is
axially symmetric and that furthermore the only radial geodesics
that can emerge are in the polar direction. In this situation, the
analysis of the visibility of the singularity is essentially the
same as that for the spherically symmetric case. We show here that
there are choices of the initial data for which the central
singularity is visible along the polar direction. We follow the
treatment of the spherically symmetric (LTB) case given in
\cite{mena+nolan}. However as with the previous section, some care
must be taken to account for the minor differences between the
present case and the spherically symmetric case.

The axially symmetric case is obtained by setting $B_1=B_2=0$ in
(\ref{nudef}), and the polar geodesic corresponds to $x=y=0$. Then
the null geodesic equations (\ref{geo-t})-(\ref{geo-int}) reduce to
\bq
\td^2-e^{2\alpha}\rd^2&=&0,\label{pgi}\\
\ddot{t}+\alpha_t\td^2&=&0,\label{pgt}\\
\ddot{r}+(\alpha^\prime+2\alpha_te^\alpha)\rd^2&=&0.\label{pgr} \eq

Our first step is to show that we can replace the affine parameter
$s$ by the coordinate $r$, and consider the projection of the
geodesic into the $r$-$t$ plane. To see this, we note that from
(\ref{pgr}), if $\rd(s_0)=0$ for some $s_0$, then $\rd(s)=0$ for
all $s$, and the geodesic reduces to a single point. So we can
assume that $\rd\neq0$ along the geodesic. Hence a polar null
geodesic that is initially outgoing $\rd(s_0)>0$ remains outgoing
for all $s$. Consequently, apart from the question of the maximal
$s-$interval of existence of the geodesic, all information
regarding the outgoing polar geodesic is contained in the single
equation \be \frac{dt}{dr}=e^\alpha=R^\prime+R\nu^\prime.
\label{t-r-polar}\ee

Along such a geodesic $\gamma$, we have
\[ \nu^\prime|_\gamma = -\frac{C^\prime}{C}.\]
In the axially symmetric case, the conditions (\ref{norm}) and
(\ref{a-con1})-(\ref{a-con3}) reduce to \bq
AC=\frac14,\label{ax-con1}\\A>0,\label{ax-con2}\\A-rA^\prime>0,\label{ax-con3}\\
A^\prime C^\prime>-\frac{1}{4r^2}.\label{ax-con4}\eq From these we
obtain \be -\frac{1}{r}<\nu^\prime|_\gamma
<\frac{1}{r}.\label{nucons}\ee Thus the additional symmetry in the
problem yields a useful additional bound on $\nu^\prime$ ({\em cf.}
(\ref{nu-con1})).

To proceed, we make an additional mild assumption on the structure
of the function $F$. We define the number $f_0$ and the function
$F_1$ by
\[ F=r^3(f_0+F_1(r)),\quad F_1(0)=0.\]
Our mild assumption is that $f_0>0$: this corresponds to the
initial central density being strictly positive. The no-shell
crossing condition corresponds to $F_1^\prime<0$ for $r>0$, and so
$f_1:=F_1^\prime(0)\leq0$.

\begin{prop}
If $f_1<0$, then there is a future pointing outgoing polar geodesic
with past endpoint on the central singularity.
\end{prop}
\noindent{\bf Proof:} For $f_1<0$, we can use (\ref{ah-def}) to
write
\[ t_{ah}=\frac23f_0^{-1/2}-\frac13f_0^{-3/2}f_1r+O(r^2),\]
where this and all other asymptotic relations in the present proof
refer to the limit $r\to 0$. For constant $\kappa$ with
$0<\kappa<-\frac13f_0^{-3/2}f_1=:\lambda$, define
\[ t_\kappa(r)=\frac23f_0^{-1/2}+\kappa r.\]
Then for sufficiently small $\delta_1$, there is a non-empty region
\[ \Omega[\delta_1,\kappa]=\{(t,r):t_\kappa(r)<t<t_{ah}(r), 0<r<\delta_1\}.\]
Note that $t_\kappa^\prime=\kappa>0$. Along a future pointing
outgoing polar geodesic $\gamma$, we have
\[ \frac{dt}{dr}=R^\prime+R\nu^\prime<R^\prime+\frac{R}{r},\]
and a straightforward calculation using (\ref{Rsol1}) then yields
\bq
\left.\frac{dt}{dr}\right|_\gamma&<&\left(\frac94\right)^{1/3}f_0^{1/3}
(\lambda-\kappa)^{2/3}\left(2-\frac{2f_1}{3f_0^{3/2}(\lambda-\kappa)}\right)~r^{2/3}\nn\\&&+O(r^{5/2}).\nn\eq
Since $f_1<0$ and $0<\kappa<\lambda$, the leading coefficient here
is positive, and so there is a $\delta_2>0$ such that for all
$r<\delta_2$, we have
\[\left.\frac{dt}{dr}\right|_\gamma<t_\kappa^\prime.\]
Hence by taking $\delta$ to be sufficiently small to allow use of
Proposition 4 and to minimize $\delta_1$ and $\delta_2$, we see
that a future-pointing outgoing polar null geodesic in the region
$\Omega[\delta,\kappa]$ as defined above cannot leave this region
as we extend back into the past. Hence the geodesic must extend
back to the central singularity $r=0,
t=t_c(0)=t_{ah}(0)=t_\kappa(0)$. \hfill$\blacksquare$

\begin{remark}
{\em We note that as in the spherically symmetric case, it is
possible to consider the case where $f_1=0$. This requires an
additional assumption on the differentiability of $F$ and on the
value of the coefficients of a Taylor expansion of the function
around $r=0$. There is nothing to indicate that the results obtained
in this way would differ from those obtained in the spherically
symmetric case.}
\end{remark}

We consider next the important question of whether or not these
geodesics meet the singularity at some finite affine parameter value
in the past, or if $s\to-\infty$ as $r\to0,t\to t_c(0)$. In order to
do this, we study the sign of $\alpha_t$ in the region
$\Omega[\delta,\kappa]$ introduced in the proof of Proposition 5
above.

\begin{lemma}
There are values of $\kappa\in(0,-\frac13f_0^{-3/2}f_1)$ and
$\delta>0$ such that $\alpha_t>0$ for all
$(t,r)\in\Omega[\delta,\kappa]$.
\end{lemma}
\noindent{\bf Proof:} We have
\[ \alpha_t=\frac{R_t^\prime+R_t\nu^\prime}{R^\prime+R\nu^\prime}\]
and so using the no-shell crossing condition $R^\prime+R\nu^\prime$,
this is positive if and only if the numerator is positive. From
(\ref{nucons}), we have
\[R_t^\prime+R_t\nu^\prime>R_t^\prime+\frac{R_t}{r}\]
(recall that $R_t<0$). The latter term is positive if and only if
\[
\frac12\left(\frac{R}{r}\right)^{-3/2}\left(\frac{1}{r}-\frac13\frac{F^\prime}{F}\right)>
\frac{1}{r}+\frac{F^\prime}{3F}=\frac{2}{r}+O(1).\] We note that
\[\frac{1}{r}-\frac13\frac{F^\prime}{F}=-\frac13\frac{f_1}{f_0}+O(r)>0.\]
For $t>t_\kappa$, we have
\[\left(\frac{R}{r}\right)^{-3/2}>\left(1-\frac{t_\kappa}{t_c}\right)^{-1}=-t_c(0)(\kappa-\lambda)^{-1}r^{-1}+O(r^2).\]
We note that the coefficient of $r^{-1}$ here is positive. Hence if
we can choose $\kappa$ so that
\[ \frac{t_c(0)}{6}\frac{f_1}{f_0}(\kappa-\lambda)^{-1}=\frac{\lambda}{3(\lambda-\kappa)}>2,\]
then there exists $\delta>0$ so that $\alpha_t>0$ on
$\Omega[\delta,\kappa]$ as required. This choice entails
$\kappa>\frac56\lambda$, which can always be
made.\hfill$\blacksquare$.

\begin{prop}
Let $\kappa$ and $\delta$ have the values required by Lemma 2. Then
any future pointing polar null geodesic that enters the region
$\Omega[\delta,\kappa]$ extends back to the central singularity in
finite affine parameter time.
\end{prop}
\noindent{\bf Proof:} It only remains to check the statement
regarding finiteness of the value of the affine parameter $s$ at
which the geodesic meets the singularity. By Lemma 2 and
(\ref{pgt}), we see that $\ddot{t}<0$ in the limit as the
singularity is approached. If the limit $t\to t_c(0)$ is reached
only as $s\to-\infty$, then we would have
\[ \lim_{s\to-\infty}\ddot{t}\geq 0,\] contradicting the statement
above. \hfill$\blacksquare$

\section{Discussion}
We have revisited the issue of the visibility of the
shell-focusing singularity in quasi-spherical dust collapse,
motivated by the observation that previous results have
incorrectly assumed that there exist radial null geodesics in
such space-times. As we have seen, this is not generally the
case: the existence of such geodesics implies an additional
symmetry of the space-time. It is worth noting that our
discussion has been restricted to the 4-dimensional case, whereas
there have been several studies carried out in higher $n+2=D\geq
5$ dimensional Szekeres space-times. We suspect that an analogous
result applies: the existence of a radial geodesic in the
space-time implies the existence of $(n-1)$ Killing fields,
leaving just $2+1$ non-ignorable coordinates. This conjecture is
based on the structure of the derivates $\pd{\nu}{x_i}$, $1\leq
i\leq n$ in the higher dimensional case. However, we have not
been able to determine the structure of the group of
transformations for the higher dimensional case that corresponds
to the transformations (\ref{flt}) that play a crucial role in
the proof of Proposition 1.

It is perhaps worth pointing out that with the benefit of
hindsight, it is not surprising to see a connection between the
existence of radial geodesics and symmetry. First, and on general
grounds, it would be unusual to see a situation in which one had a
conserved quantity (the values of the angles in this case) without
the presence of some form of symmetry. Secondly and more
specifically for this quasi-spherical situation, it is hard to
envisage a geodesic emerging orthogonally from one 2-sphere and
remaining orthogonal to other 2-spheres that it meets - unless the
centres of those 2-spheres are aligned, with the alignment
direction forming an axis of symmetry of the spacetime. This is
exactly what we see happening in Proposition 1.

In the axially symmetric case, there is one direction along which
radial geodesics exist: the polar direction. We have looked
briefly at the issue of the visibility of the singularity along
this direction in the marginally bound case, and find no
difference between the present case and the spherically symmetric
case. However, we cannot generalize our finding to either
non-radial geodesics in the axially symmetric case, or to
geodesics in the general case. The fundamental difficulty in
doing so is that one {\em cannot} project the geodesic onto the
$r$-$t$ plane and retain all information required. The geodesic
equations in the general case (no symmetry and hence no radial
assumption allowed) form a second order nonlinear dynamical
system with singular coefficients. The presence of the singular
coefficients - which correspond to the points of space-time we
are interested in analyzing - mean that standard methods of smooth
dynamical systems do not offer means of approaching the problem.
One possible approach is to rescale the dynamical system by
multiplying through by the most strongly vanishing denominator in
the singular coefficients. One then absorbs this coefficient into
a rescaled affine parameter, to obtain a smooth system. When this
is done carefully, it is possible to convert the singular point
of the original system to a stationary point of the rescaled
system. However, this procedure typically yields non-hyperbolic
equilibrium points and spurious equilibrium sets requiring the
use of centre manifold analysis. Nonetheless, it has yielded
useful results in a different context where similar problems
(singular dynamical systems) arise \cite{louise}. It would be of
interest to see if this approach could be used to study general
geodesics in the non-axially symmetric Szekeres space-time, or
indeed the non-radial (non-polar) geodesics of the axially
symmetric Szekeres space-time. The existence of additional bounds
on the metric functions (like (\ref{nucons})) would be of
considerable use in this case.

\section*{Acknowledgement}

One of the authors (UD) is grateful to TEQIP, BESU, India for
financial support. UD is thankful to CSIR, Govt. of India for
providing research project grant (No. 25(0153)/06/EMR-II). Also
UD is thankful to School of Mathematical Sciences, DCU, Ireland
for partial hospitality where the work was carried out.



\begin{thebibliography}{szekeres}
\bibitem[1]{christo1} Christodoulou D 1999 {\em Ann. Math.} {\bf 149},
183.
\bibitem[2]{yod1} Yodzis P, Seifert H-J and M\"uller zum Hagen H
1973 {\em Commun. Math. Phys.} {\bf 34} 135.
\bibitem[3]{yod2}M\"uller
zum Hagen H, Yodzis P and Seifert H-J 1974 {\em Commun. Math. Phys.}
{\bf 37} 29.
\bibitem[4]{christo2}
Christodoulou D 1984 {\em Commun. Math. Phys.} {\bf 93} 171
\bibitem[5]{gundlach1} Gundlach C 1999 {\em Living Rev. Rel.} {\bf 2}, 4.
\bibitem[6]{joshi-krolak}
Joshi P S and Krolak A 1996 {\em Class. Quant. Grav.} {\bf 13}
3069.
\bibitem[7]{deshingkar-et-al-98}
Deshingkar S S, Jhingan S and Joshi P S 1998 {\em Gen. Rel. Grav.}
{\bf 30} 1477.
\bibitem[8]{goncalves-01}
Goncalves S M C V 2001 {\em Class. Quant. Grav.} {\bf 18} 4517.
\bibitem[9]{debnath1}
Debnath U, Chakraborty S and Barrow J D 2004 {\em Gen. Rel. Grav.}
{\bf 36} 231.
\bibitem[10]{debnath2}
Debnath U and Chakraborty S 2004 {\em J. Cosmol. Astropart. Phys.}
JCAP05(2004)001.
\bibitem[11]{debnath3}
Debnath U and Chakraborty S 2005 {\em Gen. Rel. Grav.} {\bf 37} 225.
\bibitem[12]{debnath4}
Chakraborty S and Debnath U 2005 {\em Mod. Phys. Lett.} {\bf A20}
1451.
\bibitem[13]{szekeres1}
Szekeres P 1975 {\em Commun. Math. Phys.} {\bf 41} 55.
\bibitem[14]{szekeres2}
Szekeres P 1975 {\em Phys. Rev.} {\bf D12} 2941.
\bibitem[15]{kras}
Krasi\'nski A 1997 {\it Inhomogeneous Cosmological Models}
(Cambridge University Press, Cambridge).
\bibitem[16]{magli1}
Giamb\'o R, Giannoni F and Magli G 2002 {\em Class. Quant. Grav.}
{\bf 19} L5.
\bibitem[17]{mena+nolan} Mena F C and Nolan B C 2001
{\em Class. Quantum Grav.} {\bf 18} 4531.
\bibitem[18]{louise}
Nolan L V 2007 {\em Cylindrically Symmetric Models of
Gravitational Collapse.} PhD thesis, Dublin City University
(unpublished). Singular dynamical systems arise in this work in
the consideration of self-similar, cylindrically symmetric dust
collapse.

\end{thebibliography}
\end{document}